\documentclass[11pt]{elsarticle}

    \long\def\comment#1{ }

  \usepackage{latexsym,bm,amsmath,amssymb,amsfonts}
  \usepackage{epsfig,graphics,graphicx}
  \usepackage{slashed}
  \usepackage{cite}
  \usepackage[latin1]{inputenc}
  \usepackage{mathrsfs}
\usepackage{mathcomp}

  \long\def\comment#1{ }

  \newcommand{\eqnum}[1]{Eq.~\eqref{#1}}

  \newcommand{\rmd}{{\rm d}}   

  \newcommand{\Lam}{\Lambda_{{\rm QCD}}}

  \newcommand{\beq}{\begin{eqnarray}}
  \newcommand{\eeq}{\end{eqnarray}}
  
 \def\simge{\mathrel{%
   \rlap{\raise 0.511ex \hbox{$>$}}{\lower 0.511ex \hbox{$\sim$}}}}
\def\simle{\mathrel{
   \rlap{\raise 0.511ex \hbox{$<$}}{\lower 0.511ex \hbox{$\sim$}}}}

\vspace{1.5cm}

\begin{document}
\begin{frontmatter}

\title{Parton saturation at strong coupling from AdS/CFT}

\author{Edmond Iancu}\ead{Edmond.Iancu@cea.fr}

\address{Institut de Physique Th\'eorique de Saclay,
 F-91191 Gif-sur-Yvette, France}

\date{\today}
\vspace{1.2cm}
\begin{abstract}
I describe the parton picture at strong coupling emerging from the
gauge/gravity duality, with emphasis on the universality of the
phenomenon of parton saturation. I discuss several consequences of this
picture for the phenomenology of a strongly coupled quark--gluon plasma,
which are potentially relevant for heavy ion collisions at RHIC and LHC.
\end{abstract}

\end{frontmatter}

\section{Introduction: Why study saturation at strong coupling ?}
\setcounter{equation}{0}

Since the idea of parton saturation has first emerged
\citep{Gribov:1984tu,Mueller:1985wy}, in the mid eighties, as a possible
solution to the unitarity problem in QCD at high energy, this phenomenon
has been generally associated with weak coupling. This was based on the
asymptotic freedom of QCD together with the following, self--consistent,
argument: in the kinematical domain for saturation, one expects parton
densities to be large, hence the relevant values of the QCD running
coupling should be weak; and indeed calculations in QCD at weak coupling
($\alpha_s\ll 1$) predict large gluon occupation numbers at saturation,
$n\sim 1/\alpha_s\gg 1$, thus closing the argument. Following this logic,
and also by lack of non--perturbative tools, all the subsequent studies
of this phenomenon from first principles were performed within
perturbative QCD, with increasingly higher degrees of sophistication
\citep{CGCreviews} (and Refs. therein). In particular, the observation
that a regime characterized by high occupation numbers and weak coupling
is semi--classical \citep{McLerran:1993ni} paved the way to the modern
effective theory for gluon saturation within pQCD, which is the Color
Glass Condensate (CGC) \citep{CGCreviews}. These studies demonstrated the
existence of an intrinsic scale associated with this phenomenon, the {\em
saturation momentum} $Q_s$  --- the transverse momentum below which
non-linear effects in the gluon distribution  become important ---, which
increases quite fast with the energy, and thus eventually becomes `hard'
($Q_s \gg \Lam \simeq 200$ MeV). This scale also controls the gluon
density at and near saturation, and hence it sets the scale for the
running coupling in the approach towards saturation. All these results
have confirmed the original intuition that, {\em for sufficiently high
energies}, parton saturation is a weak coupling phenomenon which is
driven by the rapid evolution of the gluon distribution via
bremsstrahlung.

But what about the {\em current} energies, as attained in the present
days colliders ? These energies are relatively high, allowing to explore
values of the Bjorken's $x$ variable --- the longitudinal momentum
fraction of a parton inside the hadron wavefunction --- as small as
$x\sim 10^{-4}$ at RHIC, $10^{-5}$ at HERA and even $10^{-6}$ at the LHC.
Besides, in collisions involving large nuclei, the gluon density and thus
the saturation momentum are further enhanced by the atomic number $A \gg
1$. Yet, in spite of such favorable circonstances, the corresponding
values of $Q_s$ remain quite modest: this scale does not exceed $1.5$~GeV
at HERA and RHIC (with nuclei) and it should be around $2\div 3$~GeV in
the `forward' kinematics at the LHC. Moreover, the theoretical analyses
of the phenomenology initiate the high--energy evolution at some
intermediate value $x_0$ which must be small enough to justify the focus
on the evolution with increasing energy (as opposed, e.g., to the DGLAP
evolution), but large enough (with respect to the $x$ values of interest)
to minimize the effects of the uncertainties in the initial conditions at
$x_0$ (which must be taken from a model, so like the
McLerran--Venugopalan model \citep{McLerran:1993ni}). This value $x_0$
and the associated saturation momentum $Q_s(x_0)$ are generally chosen in
such a way to optimize the description of some set of data, so like the
HERA data for the DIS structure function $F_2$, and some typical values
(taken from Ref. \citep{Albacete:2009fh}) are $x_0\sim 0.01$ and
$Q_{s}^2(x_0)\,\sim\, 0.4$~GeV$^2$. For such `semihard' values $Q_s$, the
weak coupling techniques are only marginally applicable.

So, clearly, it would be very interesting to have some (at least,
qualitative) understanding of the phenomenon of parton saturation in the
transition region towards strong coupling. In general, that problem is
extremely complicated not only because the coupling is strong, but
especially because of the possible mixing with the physics of
confinement, for which there is no analytic understanding from first
principles. It is therefore both interesting and remarkable that there
exists a physical regime of QCD, in which one can isolate the physics of
(relatively) strong coupling from that of confinement, and which moreover
might have some relevance for the present day phenomenology, as suggested
by some of the data at RHIC. This refers to the deconfined phase of QCD,
the {\em quark--gluon plasma} (QGP), which in thermodynamical equilibrium
exists for temperatures larger than a critical value $T_c\simeq 170$~MeV.
This phase has been rather extensively studied (at least, in so far as
its thermodynamical properties are concerned) via lattice QCD
calculations. By now we have rather firm evidence that it has been also
experimentally produced at RHIC, in the intermediate stages of the heavy
ion collisions \citep{Gyulassy:2004zy}. There are moreover strong
indications that the partonic matter liberated by the collision
equilibrates quite fast, over a time $\tau_0\sim 1$~fm/c, at a
temperature $T=(2\div 3) T_c$, and then lives in the plasma phase for
about 5~fm/c, before eventually cooling down and hadronizing. (In
lead--lead collisions at LHC, one should reach $T\sim 5 T_c$ and a QGP
lifetime $\tau\sim 10$~fm/c.)

In the forthcoming two sections, I shall argue that: \texttt{(i)} there
are strong indications, notably from the heavy ion experiments at RHIC,
that this deconfined matter is effectively strongly coupled,
\texttt{(ii)} the physics of parton saturation in a strongly coupled
plasma is interesting not only at a conceptual level, but also for the
phenomenology at RHIC, and \texttt{(iii)} this physics can be reliably
studied, at least at a qualitative level, by using the gauge/gravity
duality. Then, in the remaining part of the discussion, I will explain
the consequences of this approach for parton evolution and saturation at
strong coupling.

\section{sQGP at RHIC}
\setcounter{equation}{0}

There are several arguments why the quark--gluon plasma produced at RHIC
or LHC can be considered as strongly coupled. At a formal level, one can
note that the relevant temperatures are relatively low (less than 1 GeV),
so the respective QCD coupling is quite high: $\alpha_s\equiv g^2/4\pi
\simeq 0.4$, or $g\simeq 2$. Moreover, unlike at zero temperature, where
the perturbative expansion is a series in powers of $\alpha_s$, at finite
temperature this is truly a series in powers of $g$ and it shows very bad
convergency unless $g\ll 1$ (which in QCD requires astronomically high
temperatures) \citep{Blaizot:2003tw}. But this formal argument is not
decisive by itself, as shown by the following fact: one has demonstrated
that appropriate resummations of the perturbative expansion are able to
cure the problem of the lack of convergency and thus yield results for
the QCD thermodynamics which agree very well with lattice QCD for all
temperatures $T\gtrsim 2.5 T_c$ \citep{Blaizot:2003tw}. Underlying such
resummation schemes, there is the picture of QGP as a gas of weakly
interacting `quasiparticles' --- quarks and gluons with energies and
momenta of order $T$ which are `dressed' by medium effects.

But this picture, which would be natural at weak coupling, has been
shaken by some of the experimental discoveries at RHIC
\citep{RHICglobal}, especially the unexpectedly large `elliptic flow' and
`jet quenching', which are rather suggestive of strong coupling
\citep{Gyulassy:2004zy,Muller:2007rs}.

The `elliptic flow' refers to an azimuthal anisotropy in the distribution
of the particles produced in a peripheral nucleus--nucleus collision.
Such a pattern is natural for a {\em fluid}, which is a system with
strong interactions, but it would be very difficult to explain for a
weakly coupled gas. The elliptic flow measured at RHIC \citep{RHICglobal}
not only is strong, but it is so even for the heavy quarks $c$ and $b$,
which appear to be dragged by the medium in spite of their large masses.
The RHIC data for elliptic flow can be well accommodated within
theoretical analyses using hydrodynamics, which assume early
thermalization ($\tau_0 \lesssim 1$~fm/c) and small viscosity --- more
precisely, a very small viscosity to entropy--density ratio $\eta/s$.
These features are signatures of a system with strong interactions:
indeed, when $g\ll 1$, both the equilibration time $\tau_0$ and the ratio
$\eta/s$ are parametrically large, since proportional to the mean free
path $\sim 1/g^4$. On the other hand, AdS/CFT calculations for gauge
theories with a gravity dual \citep{Policastro:2001yc} suggest that, in
the limit of an infinitely strong coupling, the ratio $\eta/s$ should
approach a universal lower bound which is $\hbar/4\pi$
\citep{Kovtun:2004de}. Interestingly, it appears that, within the error
bars, the ratio $\eta/s$ extracted (via the theoretical analyses) from
the RHIC data \citep{Luzum:2008cw} is rather close to this lower bound,
thus supporting the new paradigm of a {\em strongly coupled Quark--Gluon
Plasma} (sQGP).

Whereas the elliptic flow is a manifestation of long--range correlations
which could be indeed sensitive to larger values of the coupling, the
observation of strong--coupling aspects in relation with `jet quenching'
looks even more surprising, since it seems to be in conflict with
asymptotic freedom. The `jet quenching' refers to the energy loss and
transverse momentum broadening of an energetic parton (the `jet') which
interacts with the medium. The jet has a relatively large transverse
momentum $k_\perp\gg T$ and hence it explores the structure of the plasma
on relatively small space--time distances $\ll 1/T$ (`hard probe'). Some
typical values at RHIC are $k_\perp\sim 2\div 20$~GeV and $T\sim
0.5$~GeV. Because of this large separation in scales, one would expect
the medium to be relatively transparent for the jets, but the
measurements at RHIC show that this is actually not the case: the medium
appears to be opaque \citep{RHICglobal}.

\begin{figure*}{\centerline{
\includegraphics[width=.6\textwidth]{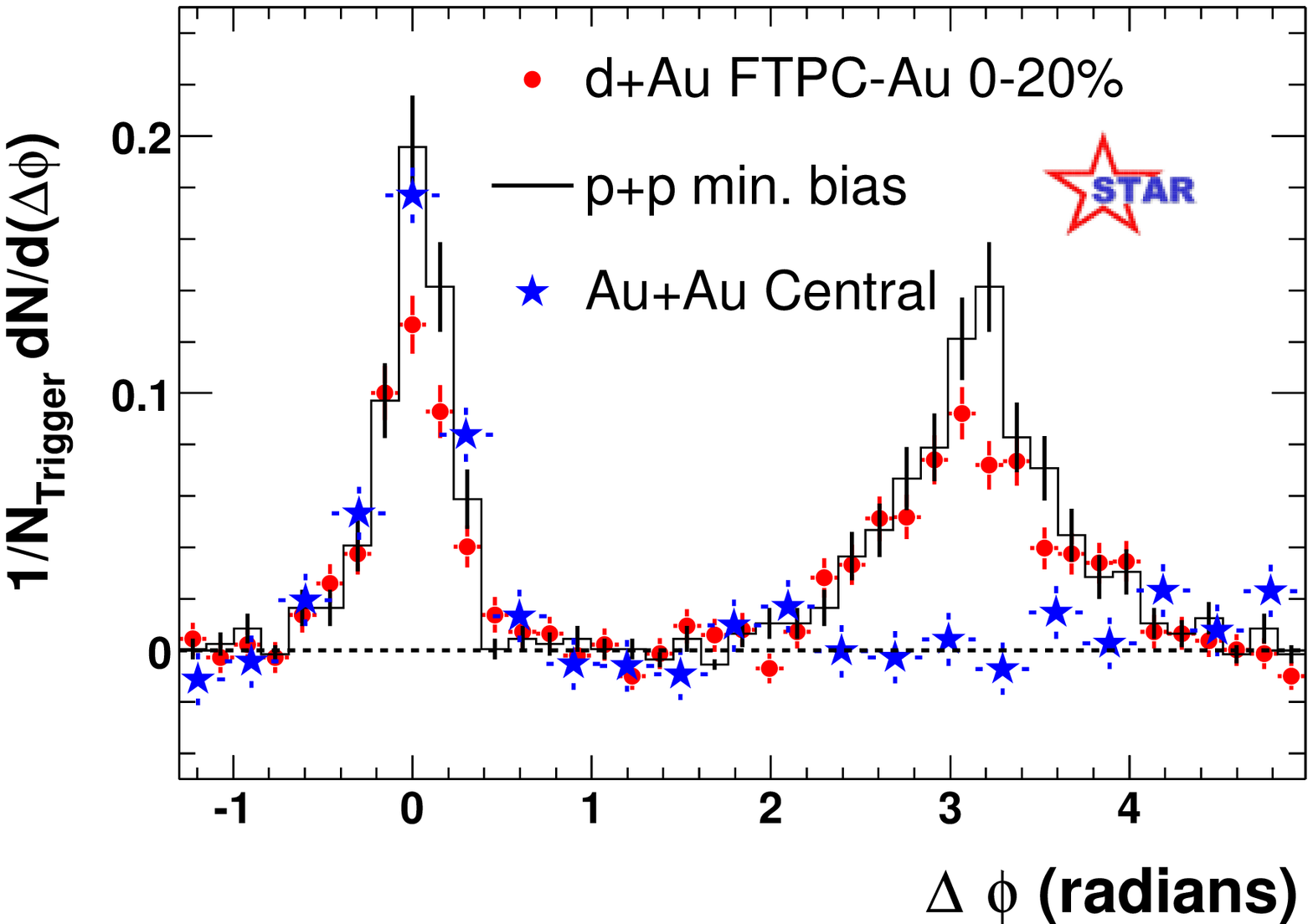}\
\includegraphics[width=.45\textwidth]{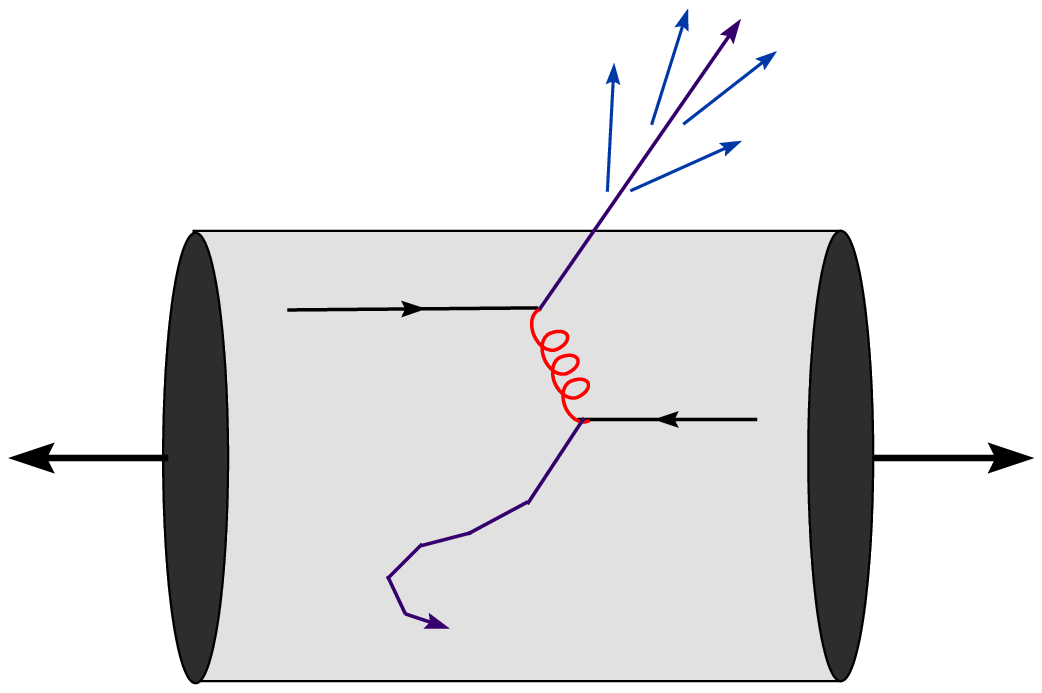}}
\caption{\sl\small Left: Azimuthal correlations for jet measurements
($k_\perp(assoc) > 2$~GeV) at RHIC (STAR) in p+p, d+Au, and Au+Au
collisions. Right: Jet production in a nucleus--nucleus collision.}
\label{fig:JETS}}
\end{figure*}

This opaqueness is manifest e.g. in the RHIC measurements of the `nuclear
modification factor' --- the ratio $R_{AA}$ between the particle yield in
Au+Au collisions and the respective yield in proton--proton collisions
scaled up by $A^2$. This ratio would be one in the absence of medium
effects, but in reality one finds a much lower value, $R_{AA}\simeq
0.2\div 0.3$, which is interpreted as a sign of strong energy loss in the
medium. Another observable which points in the same direction is the
`away--jet suppression' observed in the azimuthal correlations of the
produced jets: unlike in p+p or d+Au collisions, where the hard particles
typically emerge from the collision region as pairs of back--to--back
jets, in the Au+Au collisions at RHIC one sees `mono--jet' events in
which the second jet is missing (see Fig.~\ref{fig:JETS} left). This has
the following natural interpretation (see Fig.~\ref{fig:JETS} right): the
hard scattering producing the jets has occurred near the edge of the
interaction region, so that one of the jets has escaped and triggered a
detector, while the other one has been deflected, or absorbed, via
interactions in the surrounding medium.


\begin{figure*}{ \centerline{
\includegraphics[width=.46\textwidth]{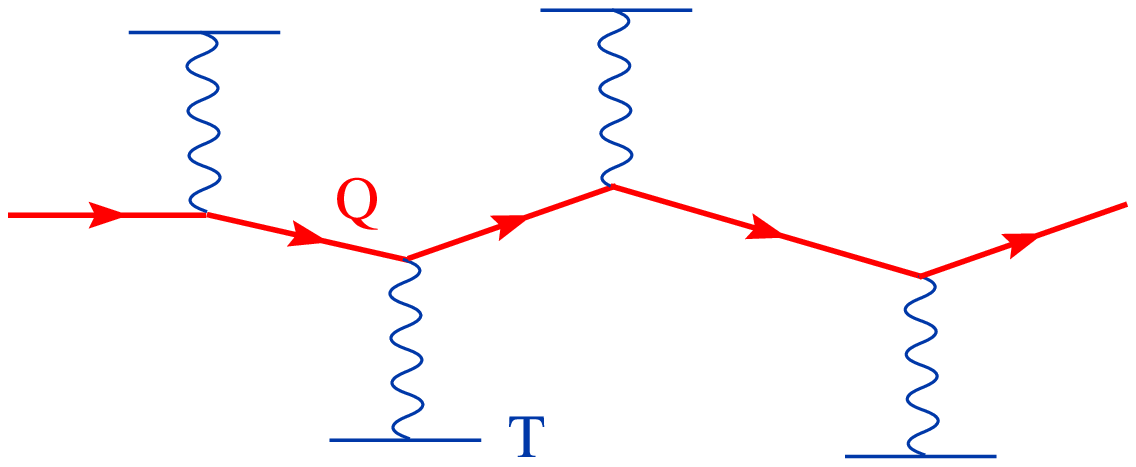}\qquad
\includegraphics[width=.46\textwidth]{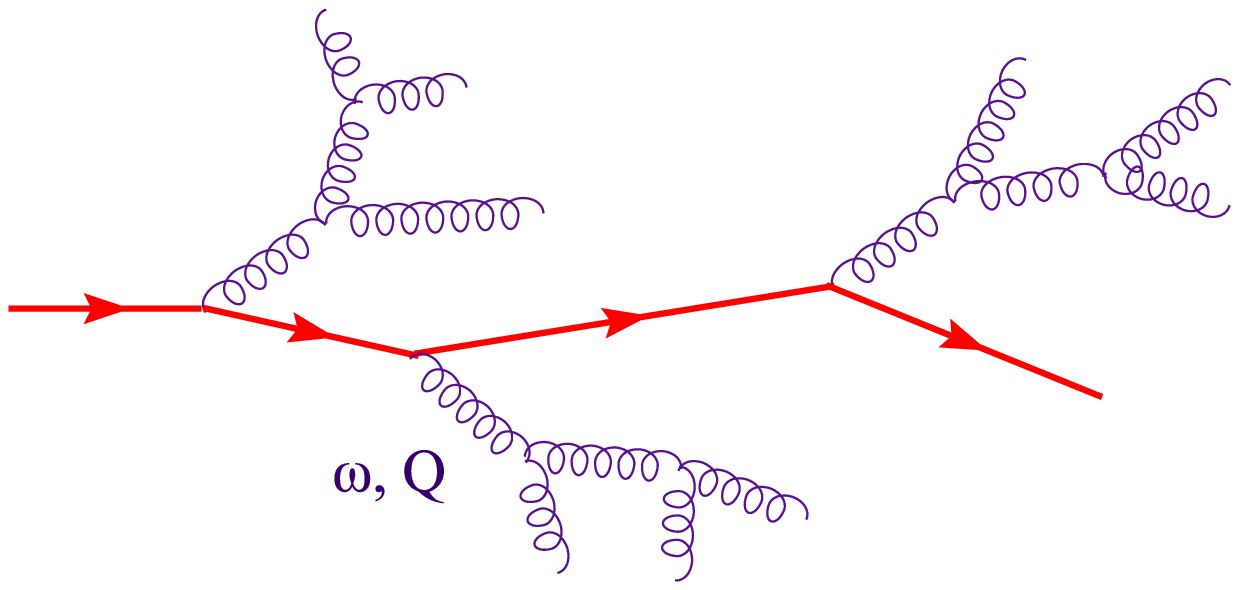}}
\caption{\sl\small Transverse momentum broadening for a heavy quark which
propagates through a quark--gluon plasma. Left: weak coupling (successive
scattering off thermal quasiparticles). Right: strong coupling (medium
induced branching).} \label{fig:quench}}
\end{figure*}

\begin{figure}[htb]{\centerline{
\includegraphics[width=.8\textwidth]{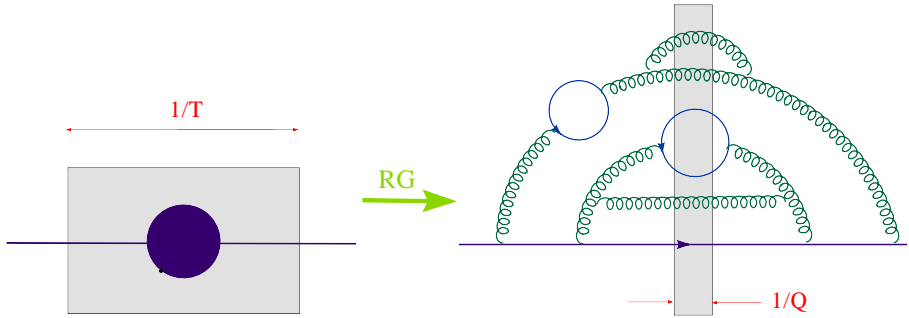}}
\caption{\small\sl Parton evolution from the thermal scale $T$ up to the
harder scale $Q\gg T$.}  \label{fig:RG}}
\end{figure}

Assuming weak coupling, it is possible to compute energy loss and
momentum broadening within perturbative QCD \citep{Baier:1996kr}. If the
medium is composed of weakly interacting quasiparticles (quarks and
gluons), then the deflection of the hard jet is due to its successive
scattering off these quasiparticles (see Fig.~\ref{fig:quench} left).
Also, energy loss at weak coupling is dominated by {\em medium induced
radiation}, that is, the emission of a hard gluon in the presence of
medium rescattering. Both phenomena are controlled by the same transport
coefficient, the `jet quenching parameter' $\hat q$, defined as the rate
of transverse momentum broadening. In pQCD $\hat q$ is estimated as the
cross--section for the scattering between the jet and the plasma
constituents `seen' by the jet on its hard resolution scale. At high
energy, these constituents are mostly gluons and $\hat q$ is estimated as
\citep{Baier:1996kr}
 \begin{eqnarray}\label{qhat}
 \hat q\,\equiv\, \frac{{\rm d} \langle k_\perp^2\rangle}{{\rm d} t}
  \,\simeq\, \frac{\alpha_s N_c}{N_c^2-1}\,
   \,{\mathcal G}(x,Q^2),
 \end{eqnarray}
where ${\mathcal G}(x,Q^2)$ is the gluon distribution in the medium on
the resolution scale $Q^2\sim \langle k_\perp^2\rangle$, as produced via
the quantum evolution of the quasiparticles from their intrinsic energy
scale to the hard scale $Q$ (see Fig.~\ref{fig:RG}). For instance, if the
medium is a finite--temperature plasma with temperature $T$, then
${\mathcal G}\simeq n_q(T)\,{\mathcal G}_q\,+\,n_g(T)\,{\mathcal G}_g$,
where $n_{q,g}(T)\propto T^3$ are the quark and gluon densities in
thermal equilibrium and ${\mathcal G}_{q,g}(x,Q^2)$ are the gluon
distributions produced by a single quark, respectively gluon, on the
scale $Q\gg T$. $\hat q$ is also related to the saturation scale $Q_s$ in
the plasma, via $Q_s^2 \simeq \hat q L$ where $L$ is the longitudinal
extent of the medium. At weak coupling, one can evaluate all these
quantities within pQCD. By doing that, one finds an estimate $\hat q_{\rm
pQCD}\simeq (0.5\,\div\,1) {\rm GeV}^2/{\rm fm}$, while phenomenology
\citep{Abelev:2006db,Adare:2006nq} rather suggests that $\hat q$ should
be somehow larger, between 5 and 15 ${\rm GeV}^2/{\rm fm}$. One should
nevertheless keep in mind that this phenomenology is quite difficult and
not devoid of ambiguities: strong assumptions are necessary in order to
compute $\hat q$, and also to extract its value from the RHIC data (see,
e.g., the discussion in \citep{Baier:2006fr}).

This discrepancy suggests that the actual gluon distribution in the
plasma is significantly larger than expected in pQCD. A possible
explanation for that is a stronger value for the coupling, which would
enhance the quantum evolution from $T$ up to $Q$. Note that there is not
necessarily a conflict with asymptotic freedom: to get an enhanced gluon
distribution on the relatively hard scale $Q$, it is enough to have a
stronger coupling at the lower end of the evolution, that is, at the
relatively soft scale $T$ (where we know that $g\simeq 2$ is indeed quite
large). Actually, in Ref. \citep{Iancu:2009zz} we proposed a strategy for
numerically studying this evolution in lattice QCD at finite temperature
and thus directly test the hypothesis of strong coupling.

\section{The AdS/CFT correspondence}

The previous discussion invites us to a better understanding of parton
evolution and saturation in deconfined QCD matter at strong coupling,
that is, for $\alpha_s\equiv g^2/4\pi \simeq 1$. However, even without
the complications of confinement, the QCD calculations at strong coupling
remain notoriously difficult. (In particular, lattice QCD cannot be used
for real--time phenomena so like scattering.) So it has become common
practice to look to the ${\mathcal N}=4$ supersymmetric Yang--Mills (SYM)
theory, whose strong coupling regime can be addressed within the AdS/CFT
correspondence, for guidance as to general properties of strongly coupled
plasmas (see the review papers
\citep{Son:2007vk,Iancu:2008sp,Gubser:2009sn}).

${\mathcal N}=4$ SYM has the `color' gauge symmetry SU$(N_c)$, so like
QCD, but differs from the latter in some other aspects: it has conformal
symmetry (the coupling $g$ is fixed) and no confinement, and all the
fields in its Lagrangian (gluons, scalars, and fermions) transform in the
adjoint representation of SU$(N_c)$. But these differences are believed
not to be essential for a study of the quark--gluon plasma phase of QCD
in the temperature range of interest for heavy ion collisions at RHIC and
LHC ($2T_c\lesssim T \lesssim 5T_c$), where QCD itself is known (e.g.,
from lattice studies \citep{Cheng:2007jq}) to be nearly conformal.

\begin{figure*}[t]
\centerline{\includegraphics[width=0.5\textwidth]{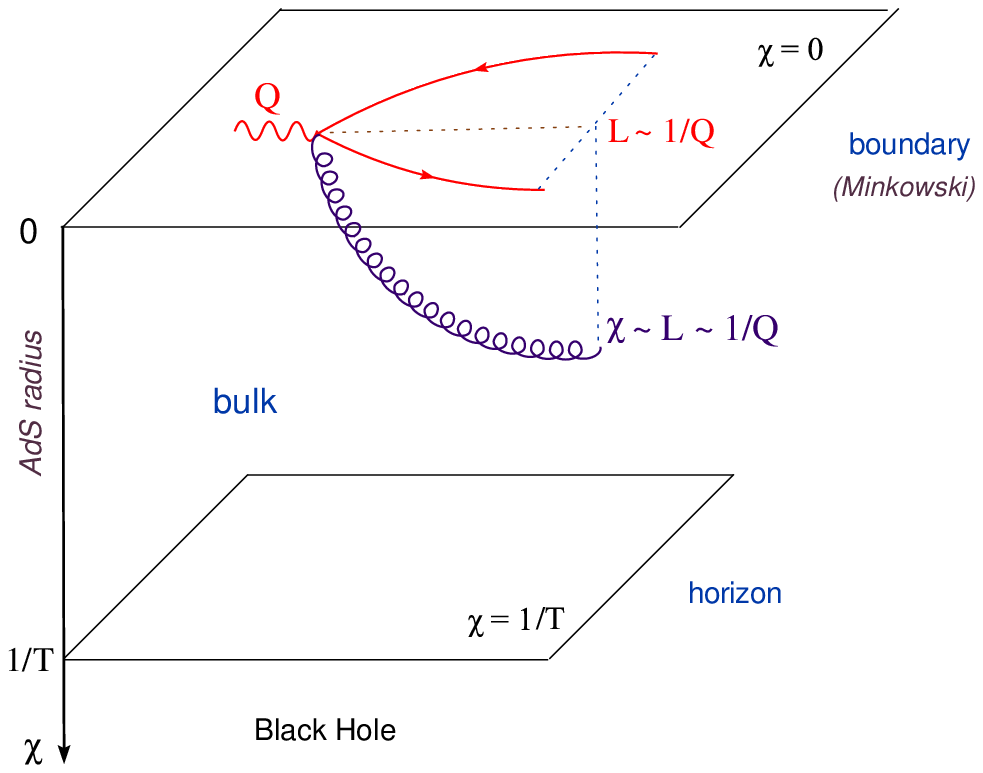} 
\includegraphics[width=0.48\textwidth]{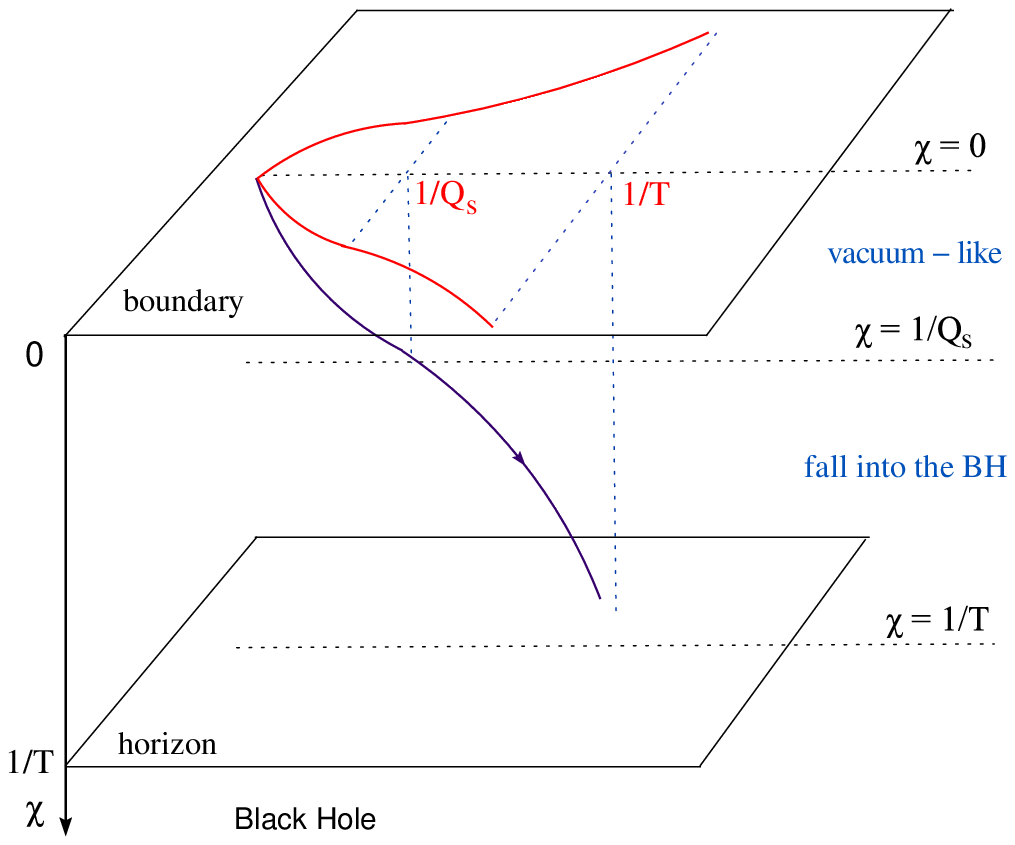}}
\caption{\small\sl  Space--like current in the plasma: the trajectory of
the wave packet in $AdS_5$ and its `shadow' on the boundary. Left: low
energy
 ---  the Maxwell wave gets stuck near the
boundary. Right: high energy --- the wave falls into the BH.
\label{fig:WAVE}}
\end{figure*}

The AdS/CFT correspondence
\citep{Maldacena:1997re,Gubser:1998bc,Witten:1998qj} is the statement
that the conformal field theory (CFT) ${\mathcal N}=4$ SYM is `dual'
({\em i.e.}, equivalent) to a specific string theory (`type IIB') living
in a $(9+1)-$dimensional space time with AdS$_5\times S^5$ geometry. The
5--dimensional Anti-de-Sitter space--time AdS$_5$ is a space with Lorentz
signature and uniform negative curvature and can be roughly imagined as
the direct product between our $(3+1)-$dimensional Minkowski world and a
radial, or `fifth', dimension $\chi$, with $0\le \chi < \infty$. Our
physical world is the boundary of AdS$_5$ at $\chi=0$ (see the sketch in
Fig.~\ref{fig:WAVE}). The radial dimension is, roughly speaking, dual to
the virtual momenta of the quantum fluctuations that we implicitly
integrate out in the boundary gauge theory (see the discussion in
Sect.~\ref{DIS}).

This gauge/string equivalence is conjectured to hold for arbitrary values
of the parameters $g$ and $N_c$, but in practice this is mostly useful in
the strong `t Hooft coupling limit $\lambda\equiv g^2N_c\to \infty$ with
$g\ll 1$, where the string theory becomes tractable --- it reduces to
classical gravity in 9+1 dimensions (`supergravity'). This limit is
generally not a good limit for studying scattering, since the respective
amplitudes are suppressed as $1/N_c^2$ \citep{Polchinski:2002jw,HIM1}.
Yet, this is meaningful for processes taking place in a deconfined
plasma, which involves $N_c^2$ degrees of freedom per unit volume, thus
yielding finite amplitudes when $N_c\to\infty$. The gravity dual of the
${\mathcal N}=4$ SYM plasma with temperature $T$ is obtained
\citep{Witten:1998zw} by introducing a black--hole (BH) in the radial
dimension of AdS$_5$ --- something that may look natural, given that a BH
has entropy and thermal (Hawking) radiation. The BH horizon is located at
$\chi=1/T$ and is parallel to the Minkowski boundary --- that is, the BH
is homogeneous in the physical 4 dimensions. One can see here a
manifestation of the {\em ultraviolet/infrared correspondence} (or
`holographic principle'), which is very useful for the physical
interpretation of the supergravity calculations: the presence of a
gravitational source at a distance $\chi_0$ in the bulk of AdS$_5$ (here
the BH horizon at $\chi_0=1/T$) corresponds to adding a energy/momentum
scale $1/\chi_0$ in the boundary gauge theory (here, the plasma with
temperature $T$).

\section{Deep inelastic scattering at strong coupling}
\label{DIS}

To understand parton evolution at strong coupling, we need a
non--perturbative and gauge--invariant definition of the concept of
`parton', which can be extrapolated to AdS/CFT. This is provided by deep
inelastic scattering (DIS), which in its essence describes the absorption
of a space--like photon by the constituents of the target hadron which
carry electric charge. The respective cross--section --- the `structure
function' $F_2(x,Q^2)$ --- is a direct mesure of the distribution of
these charged partons (the quarks in the case of QCD) on the resolution
scales $x$ and $Q^2$ of the virtual photon. Here $Q^2$ is the photon
virtuality and fixes the typical transverse momentum (or inverse
transverse size) of the struck quark. Furthermore $x\equiv Q^2/(2q\cdot
P)\approx Q^2/s$, with $s$ the invariant energy squared of the
proton--photon system, is the Bjorken--$x$ variable and fixes the
longitudinal resolution of the photon: the struck quark carry a fraction
$x$ of the hadron longitudinal momentum $P$. At weak coupling at least,
the gluon distribution can be extracted too from the measured
$F_2(x,Q^2)$, by using the perturbative evolution equations for the
parton distributions.

Using the AdS/CFT correspondence we have computed DIS off the ${\mathcal
N}=4$ SYM plasma at temperature $T$ and in the strong coupling (or
large--$N_c$) limit \citep{HIM2,HIM3}. The virtual photon couples to the
constituents of the plasma which carry the ${\mathcal R}$--charge (the
analog of the electromagnetic charge in ${\mathcal N}=4$ SYM). The dual,
supergravity, picture of DIS is as follows (see Fig.~\ref{fig:WAVE}) :
the ${\mathcal R}$--current $J_\mu$ acts as a perturbation on the
Minkowski boundary of AdS$_5$ at $\chi=0$, thus inducing a massless,
vector, supergravity field $A_m$ (with $m=\mu$ or $\chi$) which
propagates towards the bulk of AdS$_5$ ($\chi>0$), according to Maxwell
equations in curved space--time\footnote{The 5--dimensional sphere
($S^5$) part of AdS$_5\times S^5$ plays no role for this particular
calculation.} :
 \begin{eqnarray}\label{maxwell} \partial_m\big(\sqrt{-g}g^{mp}g^{nq}
 F_{pq})\,=\,0\,,
 \qquad\mbox{where}\quad F_{mn}=\partial_m A_n-\partial_n A_m\,.
 \end{eqnarray}
These equations describe the gravitational interaction between the
Maxwell field $A_m$ and the BH (implicit in the 5--dimensional metric
tensor $g^{mn}$). Note that there is no explicit coupling constant in the
equations: the gravitational scattering is rather controlled by the
kinematics. Given the solution $A_m$, there is a well--identified
procedure to construct the current--current correlator $\langle J_\mu(x)
J_\nu(y)\rangle$ in the boundary gauge theory. Then, the DIS structure
function is finally obtained by taking the imaginary part of this
correlator in momentum space, like usual.

Eqs.~(\ref{maxwell}) remain non--trivial even at $T=0$, in which case
they describe the propagation of the virtual photon through the vacuum of
the strongly--coupled ${\mathcal N}=4$ SYM theory. The physical
interpretation of the results can be deduced using the UV/IR
correspondance alluded to above, which is more precisely formulated as
follows \citep{HIM2,HIM3}: the radial penetration $\chi$ of the Maxwell
field $A_m$ in $AdS_5$ is proportional to the transverse size $L$ of the
typical quantum fluctuations of the virtual photon in the boundary gauge
theory.

As an example, consider the $T=0$ case: as well known, a space--like
photon cannot decay into on--shell (massless) quanta in the vacuum,
because of energy--momentum conservation. Rather it fluctuates into a
virtual system of partons, whose complexity can be very high at strong
coupling, but whose space--time delocalization is fixed by the
uncertainty principle: in a frame where the photon has 4--momentum
$q^\mu=(\omega,0,0,q)$ and (space--like) virtuality $Q^2= q^2-\omega^2 >
0$, its virtual fluctuations have a typical transverse size $L\sim 1/Q$
and a longitudinal size, or lifetime, $\Delta t\sim \omega/Q^2$. And
indeed, the solutions to Eqs.~(\ref{maxwell}) at $T=0$ show that the
wave--packet representing $A_m$ penetrates into the bulk up to a maximal
distance $\chi\sim 1/Q$ and the propagation time from the boundary up to
that maximal distance is $\sim \omega/Q^2$ \citep{HIM3}. The requirement
of energy--momentum conservation enters the AdS calculation in the form
of a repulsive barrier around $\chi\sim 1/Q$ which prevents the wave
packet to penetrate further down into AdS$_5$.

The same repulsive barrier, with a height $\propto Q^2$, shows up also at
finite temperature\footnote{The  photon has 4--momentum
$q^\mu=(\omega,0,0,q)$ in the plasma rest frame and we assume that
$Q^2\equiv  q^2-\omega^2\gg T^2$, as appropriate for hard probes.}, but
in that case there is also an attractive interaction $\propto
\omega^2T^4$, namely the gravitational attraction by the BH. The physical
interpretation of the latter in the dual gauge theory is quite subtle
\citep{HIM2} and will emerge from the arguments below. The competition
between these two interactions depends upon the kinematics. For
sufficiently low energy $\omega$, such that $\omega T^2\ll Q^3$, the
repulsive barrier wins and then the Maxwell field is stuck within a
distance $\chi\lesssim 1/Q\ll 1/T$ from the Minkowski boundary, so like
in the vacuum (cf. Fig.~\ref{fig:WAVE} left). In this regime there is
essentially no interaction with the black hole, meaning no absorption of
the virtual photon by the plasma, and hence no DIS. But for higher
energies and/or temperatures, such that $\omega T^2\gtrsim Q^3$, the
attraction wins and then the wave--packet falls into the BH horizon, from
which it cannot escape back anymore (cf. Fig.~\ref{fig:WAVE} right): the
space--like photon is completely absorbed into the plasma.

To understand the existence of these two regimes, it is useful to note
that the `criticality' condition for having strong interactions, that is
$\omega T^2\sim Q^3$, can be rewritten as
 \begin{eqnarray}\label{Qs}
 Q\,\sim\,\frac{\omega}{Q^2}\ T^2\,, 
 \end{eqnarray}
with the following interpretation \citep{HIM2} : the scattering becomes
strong when the lifetime $\Delta t\sim {\omega}/{Q^2}$ of the partonic
fluctuation is large enough for the mechanical work $W=\Delta t\times
F_T$ done by the {\em plasma force} $F_T\sim T^2$ acting on these partons
to compensate for their energy deficit $\sim Q$. This mechanical work
allows the partons to become (nearly) on--shell --- or more precisely to
reduce their virtuality from the original value $Q\gg T$ down to a value
of order $T$. When this happens, the fluctuation {\em thermalizes} ---
the partons become a part of the thermal bath --- and the photon
disappears into the plasma.

This plasma force $F_T\sim T^2$ represents (in some average way) the
effect of the strongly--coupled plasma on partonic fluctuations and can
be viewed as a prediction of the AdS/CFT calculation. Note that one
cannot interpret this force in terms of individual collisions between the
virtual photon and some `plasma constituents' : the BH dual to the plasma
is homogeneous in the four physical dimensions, hence it cannot transfer
any 4--momentum to the photon, so like a genuine scattering would do
(recall, e.g., Fig.~\ref{fig:quench} left). Rather, this is a kind of
{\em tidal force} which pulls the partons apart until they disappear in
the plasma. The emergence of a tidal force, which is a hallmark of
gravitational interactions, in the context of a gauge theory may look
surprising, but in fact this is not more mysterious than the basic
paradigm of the gauge/gravity duality --- the fact that a gauge theory at
strong coupling can be effectively described as gravity. Further insight
in that sense comes from an argument \citep{Polchinski:2002jw} based on
the operator product expansion (OPE): among the infinitely many
leading--twist operators which {\em a priori} contribute to OPE for DIS,
there is only one which survives in the strong coupling limit --- the
energy--momentum tensor $T_{\mu\nu}$. All the other operators acquire
large, negative, anomalous dimensions $\propto\lambda^{1/4}$ and thus are
strongly suppressed when $\lambda\to\infty$. (See also the discussion in
Sect.~\ref{SAT}.) Accordingly, one expects the theory of scattering in a
gauge theory at strong coupling to be an effective theory for
$T_{\mu\nu}$. By covariance, this must be a gravity theory.

\section{Parton saturation at strong coupling}
\label{SAT}

The OPE argument alluded to above also helps clarifying the {\em partonic
picture} of the AdS/CFT results for DIS at strong coupling
\citep{HIM2,HIM3}, to which I now turn. To formulate this picture, one
needs to use the DIS variables $Q^2=q^2-\omega^2$ and $x= Q^2/(2\omega
T)$. As previously mentioned, the AdS calculation allows one to deduce
the DIS structure function $F_2(x,Q^2)$ from the imaginary part of the
current--current correlator. The discussion in the previous section
suggests that there should be a dramatic change in $F_2(x,Q^2)$ at the
critical kinematics defined by the condition in \eqnum{Qs}. The latter
can be rewritten as $Q_s(\omega)\simeq (\omega T^2)^{1/3}$ or, in terms
of the DIS variables,
 \beq\label{xs}
 Q_s(x)\,\simeq\,\frac{T}{x}\,,\qquad\mbox{or}\qquad x_s(Q)
 \,\simeq\,\frac{T}{Q}\,.\eeq
Any of these equations defines a line in the kinematical plane
$(x,\,Q^2)$, which for reasons to shortly become clear is dubbed the {\em
saturation line}. The AdS results for DIS off the strongly coupled plasma
can then be summarized as follows \citep{HIM2} :

\texttt{(i)} For relatively low energy, or high $Q^2$, such that $x >
x_s(Q)$, the scattering is negligible and $F_2(x,Q^2)\approx 0$. (More
precisely, there is a small contribution to $F_2$ produced via tunneling
across the repulsive barrier, but this is exponentially suppressed.)

\texttt{(ii)} For higher energies, or lower $Q^2$, such that $x \lesssim
x_s(Q)$, the scattering is strong and the structure function is non--zero
and parametrically large: $F_{2}(x,Q^2)\sim x{N^2_cQ^2}$.

These results are consistent with the energy--momentum sum--rule, which
requires the integral $\int_0^1{\rm d} x\,F_2(x,Q^2)$ to have a finite
limit as $Q^2\to\infty$. (This is simply the statement that the total
energy per unit length in the plasma is the same whatever is the
resolution scale $Q^2$ on which one measures this energy: by varying
$Q^2$ one merely changes the nature and size of the partons which carry
that energy, cf. Fig.~\ref{fig:RG}, but their cumulated energy remains
the same.) Using the above results, one finds indeed
 \begin{eqnarray} \int_0^1{\rm d}
 x\,F_2(x,Q^2)\,\simeq\,\int_0^{x_s}{\rm d}
 x\,F_2(x,Q^2)\,\simeq\,
 \,x_sF_2(x_s,Q^2)\,\sim\, N_c^2 T^2
 \,, \label{SRT} \end{eqnarray}
where the integral is dominated by values $x\simeq x_s(Q)$.

The physical interpretation of these results becomes transparent after
recalling that the variable $x$ represents the longitudinal momentum
fraction of the plasma constituent which absorbs the virtual photon. Then
the above statement \texttt{(i)} implies that there are no partons at
large $x$, or high $Q^2$ : the strongly coupled plasma has no point--like
constituents. Also, statement \texttt{(ii)} together with \eqnum{SRT}
show that the total energy of the plasma as measured on a hard resolution
scale $Q^2\gg T^2$ is carried by very soft constituents with small values
of $x\simeq x_s(Q)\ll 1$.

This picture at strong coupling is very different from that of an
energetic hadron in QCD, as predicted by perturbative QCD and confirmed
by many experimental data \citep{CGCreviews}. In that case, the hadronic
wavefunction at high energy is dominated by small--$x$ partons (mostly
gluons), as produced via bremsstrahlung from partons with larger $x$.
Yet, the hadron energy is concentrated in the few partons with larger
values of $x$ (the `valence partons'); that is, the energy--momentum sum
rule is saturated by $x\sim 0.3$. Moreover these valence partons are seen
on all scales of $Q^2$, that is, there are {\em point--like}.

This rises the following questions: why and how did partons disappear at
strong coupling ? And what is the nature of the small--$x$ constituents
which carry the plasma energy on the hard resolution scale $Q^2$ ? A
first hint in that sense comes again from the OPE for DIS
\citep{Polchinski:2002jw}. The twist--two operators which enters OPE
probe the distribution of energy among the partons inside the hadron: the
hadron expectation value of the spin--$n$ twist--2 operator
$\mathcal{O}^{(n)}$ is proportional to the $(n-1)$--th moment of the
longitudinal momentum fraction $x$ carried by the quark and gluon
constituents of that hadron:
 \beq\label{On}
 \langle x^{n-1} \rangle_{Q^2}\,\equiv\,\int_0^1\rmd x\,x^{n-2}\,
 F_2(x,Q^2)\,\propto\,
 \langle\mathcal{O}^{(n)}\rangle_{Q^2}\,.
 \eeq
As indicated in this equation, the operators depend upon the resolution
scale $Q^2$, because of the quantum evolution illustrated in
Fig.~\ref{fig:RG}. In a conformal theory at strong coupling, all such
operators except for $T_{\mu\nu}$ are strongly suppressed at large
$Q^2$~: $\mathcal{O}^{(n)}(Q^2)\propto (T^2/{Q^2})^{\lambda^{1/4}}\,\to
0$. Via \eqnum{On} this implies $\langle x^{n-1} \rangle_{Q^2}\to 0$ for
any $n>2$, which suggest a very rapid evolution towards $x=0$. This is in
fact natural at strong coupling, where one expects a  {\em very efficient
parton branching}. Unlike at weak coupling, where parton radiation is
suppressed by powers of the coupling and thus favors the emission of soft
(small $x$) and collinear (small $k_\perp$) gluons --- for which the
bremsstrahlung probability is kinematically large ---, at strong coupling
there is no such a suppression anymore, and phase--space considerations
alone favor a `quasi--democratic' branching \citep{HIM2,HIM3} : the
energy and momentum of the parent parton are almost equally divided among
the daughter partons. (Soft and collinear emissions play no special role
at strong coupling, since they happen very slowly.) Via successive
branchings, all the parton will rapidly fall to small values of $x$ ---
actually the smallest one which are still consistent with
energy--momentum conservation (in the sense of the sum rule \eqref{SRT}).
That is, at strong coupling partons should still exist, but their
distributions should be concentrated at very small values of $x$. This
picture is indeed consistent with our previous findings for the strongly
coupled plasma.

One may furthermore wonder what is the mechanism which is responsible for
stopping the parton branching at sufficiently small values of $x$ and
which determines the specific $Q$--dependence of the critical value
$x_s(Q)$ --- or, vice--versa, the specific $x$--dependence of the
`saturation momentum' $Q_s(x)$. In Ref.~\citep{HIM2} we have proposed
that this mechanism is {\em parton saturation} : the partons keep
branching until their phase--space occupation numbers --- the number of
partons of a given color per unit transverse phase space $(b_\perp,
k_\perp)$ and unit rapidity $Y\equiv \ln (1/x)$ --- become of
$\mathcal{O}(1)$ :
 \beq
 \frac{1}{N^2_c}\, \frac{\rmd N}{\rmd Y \rmd^2b_\perp \rmd^2k_\perp}
 \,\simeq\,1\qquad\mbox{for}\qquad k_\perp\lesssim Q_s(x)\,=\,
 \frac{T}{x}\,. \label{phisat} \eeq
This interpretation follows from the AdS/CFT result for the
$F_{2}(x,Q^2)$ at $Q\lesssim Q_s(x)$, as shown before, together with the
observation that the quantity $(1/x)F_{2}(x,Q^2)$ is essentially the
number of partons in the plasma per unit area per unit rapidity as `seen'
by a virtual photon with resolution $Q$ :
  \beq
 \frac{1}{x}\,F_2(x,Q^2)\,\simeq\,\int^{Q}\!\! \rmd^2k_\perp\
\frac{\rmd N}{\rmd Y \rmd^2b_\perp \rmd^2k_\perp}
 \,\sim \,N_c^2Q^2\qquad\mbox{for}\qquad
Q\lesssim Q_s(x)\,.\label{nsat}
 \eeq

This interpretation is appealing in that it suggests some continuity in
the physics of saturation and unitarization from weak to strong coupling:
saturation occurs when the occupation numbers are high enough for the
repulsive interactions among the partons to prevent further radiation
\citep{CGCreviews}. At weak coupling, this requires large gluon
occupation numbers, of order $1/\alpha_s$, in order to compensate for the
weakness of the repulsive interactions. But at strong coupling, one can
think of the individual partons as `hard disks' with transverse area
$1/k_\perp^2$; then saturation occurs when these disks start to touch
with each other, {\em i.e.} for occupation numbers of order one.

But, clearly, there are important differences between the parton picture
at weak and respectively strong coupling. These differences are most
striking outside the saturation region, at $k_\perp\gg Q_s(x)$, where at
strong coupling there are no partons at all, whereas at weak coupling the
parton distributions show large `leading--twist' tails, which in fact
dominate the phenomenology at both HERA and LHC. Another important
difference refers to the energy dependence of the saturation momentum
$Q_s$. At weak coupling, this is determined by the rate for gluon
emission via bremsstrahlung, and more precisely by the BFKL evolution.
This predicts $Q_s^2 \sim 1/x^{\omega}$ where the `BFKL intercept'
$\omega$ is parametrically of $\mathcal{O}(\alpha_s N_c)$ and numerically
$\omega\simeq 0.2\div 0.3$ --- in agreement with the HERA data
\citep{CGCreviews}. At strong coupling we have found a much faster
increase with $1/x$, namely $Q_s^2(x) \propto 1/x^2$, but one factor
$1/x$ out of this result is simply a kinematical effect, related to our
study of an {\em infinite} plasma. That is, one should understand the
above result as $Q_s^2(x) \propto \Delta t/x$, where $\Delta t\sim
{\omega}/{Q^2}\sim 1/xT$ is the lifetime of the partonic fluctuation:
since the medium is infinite, the effects of the interactions accumulate
all the way along the parton lifetime. As for the other factor $1/x$,
this exhibits the `graviton intercept' $j-1=1$ ($j=2$ is the spin of the
graviton) and reflects the fact that the interactions responsible for DIS
at strong coupling involve exchanges of the energy--momentum tensor ---
the only operator which survives in OPE at strong coupling.

The above discussion also suggests how our result for $Q_s^2(x)$ should
change when, instead of an infinite plasma, we consider a slice of the
plasma with longitudinal size $L_z \ll {\omega}/{Q^2}$. In that case one
expects
 \beq\label{QsatTL}
 Q_s^2(x, T,L_z)\,\sim\, \frac{T^3 L_z}{x}\qquad
  \mbox{(slice of the plasma with $L_z\ll 1/xT$)}\,,\eeq
and this is indeed confirmed by the respective AdS/CFT calculation
\citep{Mueller:2008bt,Avsar:2009xf}. Such a `slice of the plasma' may be
viewed as a rough model for a `nucleus' in ${\mathcal N}=4$ SYM (which
however involves $N_c^2$ degrees of freedom per unit volume)
\citep{Albacete:2008vs,Albacete:2009ji,Dominguez:2008vd,Gubser:2008pc}.

\section{High--energy scattering at strong coupling}

The previously described parton picture implies that high--energy
processes taking place in the {\em vacuum} of an hypothetical world which
is conformal and strongly coupled would look quite different from the
corresponding processes in QCD. For instance, the absence of large--$x$
partons means that, in the collision between two strongly coupled
hadrons, there is no particle production at forward and backward
rapidities. This is in sharp contrast to the situation at RHIC, where the
large--$x$ partons from the incoming nuclei are seen to emerge from the
collision, as hadronic jets, along their original trajectories.

\begin{figure*}[htb]\centerline{
\includegraphics[width=.8\textwidth]{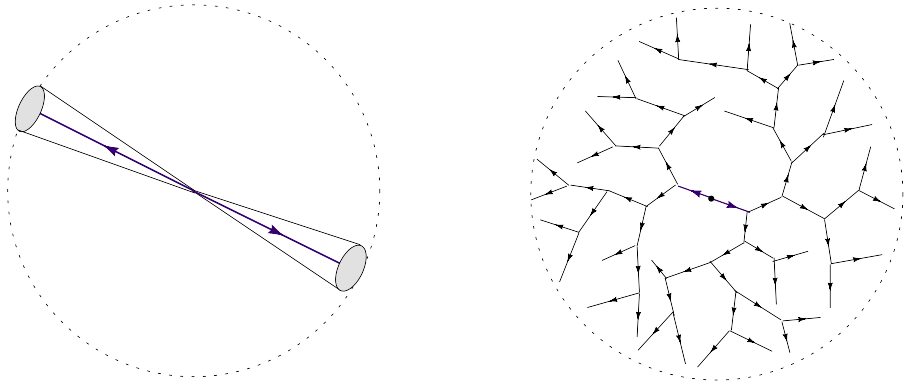}}
\caption{\sl\small $e^+e^-$ annihilation. Left: weak coupling. Right:
strong coupling. }\label{Fig:ISOTROPY}
\end{figure*}

A related prediction of AdS/CFT is the absence of jets in
electron--positron annihilation at strong coupling
\citep{HIM2,Hofman:2008ar}. Fig.~\ref{Fig:ISOTROPY} exhibits the typical,
2--jet, final state in $e^+e^-$ annihilation at weak coupling (left)
together with what should be the corresponding state at strong coupling
(right). In both cases, the final state is produced via the decay of a
time--like photon into a pair of partons and the subsequent evolution of
this pair. At weak coupling this evolution typically involves the
emission of soft and collinear gluons, with the result that the leading
partons get dressed into a pair of well--collimated jets of hadrons (cf.
Fig.~\ref{Fig:ISOTROPY} left). At strong coupling, parton branching is
much more efficient, as previously explained, and rapidly leads to a
system of numerous and relatively soft quanta, with energies and momenta
of the order of the confinement scale, which are isotropically
distributed in space (cf. Fig.~\ref{Fig:ISOTROPY} right)
\citep{Hofman:2008ar}.

But such discrepancies should not come as a surprise: after all, we know
that hard processes in high energy QCD are governed by weak coupling and
perturbative QCD does a good deal in predicting the respective
cross--sections. As for the softer processes in the vacuum, so like
hadronization, for which $\alpha_s\sim 1$, these are largely controlled
by confinement and hence they remain out of the reach of the AdS/CFT
techniques. Remember however that our original motivation for studying
high--energy processes at strong coupling is rather related to the {\em
quark--gluon plasma}, where there is no confinement and the coupling
might indeed be quite strong.

So let me finally return to this topic and consider the propagation of a
`hard probe' --- say, a heavy quark --- through a strongly--coupled
plasma. The string theory dual of a heavy quark is a Nambu--Goto string
hanging down from the boundary of AdS$_5$ and propagating through the
AdS$_5$ black hole space--time. By solving the corresponding equations of
motion, it has been possible to compute both the quark energy loss
\citep{Herzog:2006gh,Gubser:2006bz} and its momentum broadening
\citep{CasalderreySolana:2006rq,CasalderreySolana:2007qw,Giecold:2009cg}
at strong coupling. The physical interpretation of the results
\citep{Dominguez:2008vd,Giecold:2009cg} turns out to be quite
interesting: unlike in perturbative QCD, the dominant mechanism at work
is not thermal rescattering (cf. Fig.~\ref{fig:quench} left), but rather
{\em medium--induced parton branching} (cf. Fig.~\ref{fig:quench} right).

The corresponding physical picture is in fact quite similar to that of
DIS, as discussed in Sect.~4. The heavy quark can emit space--like quanta
of ${\mathcal N}=4$ SYM, which in the vacuum would have only a finite
lifetime $\Delta t\simeq \omega/Q^2$; after that time, they would be
reabsorbed by the heavy quark. (As usual, $\omega$ and $Q$ refers to the
energy and the virtuality of the emitted quanta.) However, in the
presence of the plasma, those quanta having a virtuality $Q$ lower than
$Q_s(\omega)$ can escape into the medium, and thus provide energy loss
and momentum broadening. Here, $Q_s(\omega)\simeq (\omega
 T^2)^{1/3}$ is the plasma saturation
momentum on the energy scale of the fluctuation, cf. \eqnum{Qs}.

Based on this physical picture, it is possible to estimate the rate for
energy loss: each emission brings in an energy loss $\Delta E\simeq
\omega$ over a time $\Delta t$, so the corresponding rate ${{\rm d}
E}/{{\rm d} t}$ is proportional to $\omega/\Delta t\simeq Q^2 \lesssim
Q_s^2(\omega)$. Hence, the energy loss is dominated by those fluctuations
having the maximal possible energy $\omega_{\rm max}$ and a virtuality
equal to the corresponding saturation momentum: $Q\simeq Q_s(\omega_{\rm
max})$. More precisely, the quantity which is limited is not the energy
$\omega$ of a quanta, but its `rapidity' $\gamma_p\equiv\omega/Q$ (the
Lorentz boost factor): this cannot exceed the rapidity
$\gamma=1/\sqrt{1-v^2}$ of the heavy quark (which is here assumed to
propagate at constant speed $v$). It is therefore appropriate to
reexpress $Q_s$ as a function of $\gamma_p$, by successively writing
$Q_s\simeq (\omega T^2)^{1/3}= (\gamma_p Q_s T^2)^{1/3} =
\sqrt{\gamma_p}\, T$. This quantity takes a maximal value $(Q_s)_{\rm
max}= \sqrt{\gamma}\, T$, thus yielding the following estimate for the
rate for energy loss (below, $Q_s\equiv (Q_s)_{\rm max}$) :
  \begin{eqnarray}\label{dEdt}
 -\,\frac{{\rm d} E}{{\rm d} t}\,\simeq\,\sqrt{\lambda}\,
 \frac{ \omega}{(\omega/Q_s^2)}
 \,\simeq\,\sqrt{\lambda}\,Q_s^2 \,\sim\,\sqrt{\lambda}\,\gamma\,T^2
 \,.\end{eqnarray}
The additional factor $\sqrt{\lambda}$ comes from the fact that, at
strong coupling, the heavy quark does not radiate just a single quanta
per time $\Delta t$, but rather a large number $\sim {\sqrt{\lambda}}$.
Eq.~\eqref{dEdt} is parametrically consistent with the respective AdS/CFT
result \citep{Herzog:2006gh,Gubser:2006bz}.

One can similarly estimate the momentum broadening: the $\sqrt{\lambda}$
quanta emitted during $\Delta t$ are uncorrelated with each other, so
they randomly modify the transverse momentum of the heavy quark, by a
typical amount $\Delta  k_\perp\sim Q_s$ per emission. Such random
changes add in quadrature, thus yielding
 \begin{eqnarray}\label{dpTdt}
 \frac{{\rm d} \langle k_\perp^2\rangle}{{\rm d} t}\,\sim\,
\frac{\sqrt{\lambda}\,Q_s^2}{(\omega/Q_s^2)} \,\sim\,
\sqrt{\lambda}\,\frac{Q_s^4}{\gamma Q_s}\,\sim\,
 \sqrt{\lambda}\,\sqrt{\gamma}\,T^3\,,\end{eqnarray}
in agreement with the explicit calculations in
Refs.~\citep{CasalderreySolana:2006rq,CasalderreySolana:2007qw,Giecold:2009cg}.
Note the strong enhancement of the medium effects at high energy, as
expressed by the Lorentz $\gamma$ factor in Eqs.~\eqref{dEdt} and
\eqref{dpTdt}: this might qualitatively explain the strong suppression of
particle production seen in Au+Au collisions at RHIC (cf. Sect.~2).

\section{Conclusions}

The AdS/CFT calculations summarized here demonstrate that, at least in a
conformal world, parton saturation is a universal phenomenon, appearing
at both weak and strong coupling, and presumably for any intermediate
value of the coupling. More generally, the concept of `parton' in
relation with high energy scattering appears to be relevant at strong
coupling as well. There are significant differences with respect to the
respective picture at weak coupling, so like the absence of point--like
constituents. These differences can be intuitively understood as
consequences of parton evolution at strong coupling. These differences
and their consequences for high--energy scattering --- which look very
different from the known phenomenology in QCD --- rule out this conformal
strong--coupling scenario as a candidate theory for high--energy
processes in the vacuum. On the other hand, such methods can give us a
hint towards understanding the deconfined QCD matter to be copiously
produced in the intermediate stages of heavy ion collisions at LHC. In
particular, they could shed light on fundamental open questions, such as
the rapid thermalization, the small entropy--to--density ratio, or the
strong jet quenching observed in relation with this matter at RHIC. Last
but not least, by combining in a unified theoretical framework concepts
and methods coming from fields as different as gravity, string theory,
quantum field theory, statistical physics, and hydrodynamics, the
gauge/string duality teaches us the unity of physics.


\providecommand{\href}[2]{#2}\begingroup\raggedright\endgroup

\end{document}